\documentclass[aps,prl,preprint,superscriptaddress,bibliography]{revtex4-1}

\usepackage{amsmath,amssymb,graphicx,bm}
\usepackage{xcolor}
\usepackage[colorlinks=true,urlcolor=blue,linkcolor=blue,citecolor=blue,bookmarks=false]{hyperref}

\graphicspath{{figures/}}

\begin{document}
\date{\today}
\title{Microwave spectroscopy of the low-temperature skyrmion state in Cu$_{2}$OSeO$_{3}$}

\author{Aisha Aqeel}\email{}
\affiliation{Physik-Department, Technische Universit\"at M\"unchen, D-85748 Garching, Germany}

\author{Jan Sahliger}\email{}
\affiliation{Physik-Department, Technische Universit\"at M\"unchen, D-85748 Garching, Germany}

\author{Takuya Taniguchi}
\affiliation{Physik-Department, Technische Universit\"at M\"unchen, D-85748 Garching, Germany}

\author{Stefan M\"andl}
\affiliation{Physik-Department, Technische Universit\"at M\"unchen, D-85748 Garching, Germany}

\author{Denis Mettus}
\affiliation{Physik-Department, Technische Universit\"at M\"unchen, D-85748 Garching, Germany}

\author{Helmuth Berger}
\affiliation{\'Ecole Polytechnique Federale de Lausanne, CH-1015 Lausanne, Switzerland}

\author{Andreas Bauer}
\affiliation{Physik-Department, Technische Universit\"at M\"unchen, D-85748 Garching, Germany}

\author{Markus Garst}
\affiliation{Institut f\"ur Theoretische Festk\"orperphysik, Karlsruhe Institute of Technology, D-76131 Karlsruhe, Germany}
\affiliation{Institute for quantum materials and technology, Karlsruhe Institute of Technology, D-76344 Eggenstein-Leopoldshafen, Germany}

\author{Christian Pfleiderer}
\affiliation{Physik-Department, Technische Universit\"at M\"unchen, D-85748 Garching, Germany}

\author{Christian H. Back}
\affiliation{Physik-Department, Technische Universit\"at M\"unchen, D-85748 Garching, Germany}
\affiliation{Munich Center for Quantum Science and Technology (MCQST), D-80799 M\"unchen, Germany}

\keywords{}

\begin{abstract}
In the cubic chiral magnet Cu$_{2}$OSeO$_{3}$ a low-temperature skyrmion state~(LTS) and a concomitant tilted conical state are observed for magnetic fields parallel to $\langle100\rangle$. In this work, we report on the dynamic resonances of these novel magnetic states. After promoting the nucleation of the LTS by means of field cycling, we apply broadband microwave spectroscopy in two experimental geometries that provide either predominantly in-plane or out-of-plane excitation. By comparing the results to linear spin-wave theory, we clearly identify resonant modes associated with the tilted conical state, the gyrational and breathing modes associated with the LTS, as well as the hybridization of the breathing mode with a dark octupole gyration mode mediated by the magnetocrystalline anisotropies. Most intriguingly, our findings suggest that under decreasing fields the hexagonal skyrmion lattice becomes unstable with respect to an oblique deformation, reflected in the formation of elongated skyrmions.
\end{abstract}

\maketitle

Magnetic skyrmions are topologically nontrivial spin whirls that are observed in a wide range of bulk materials, such as cubic chiral magnets, lacunar spinels, Heusler compounds, or frustrated magnets~\cite{2009_Muhlbauer_Science, 2010_Yu_Nature, 2015_Tokunaga_NatCommun, 2015_Kezsmarki_NatMater, 2017_Nayak_Nature, 2019_Kurumaji_Science}. As an immediate consequence of their nontrivial topology, the creation or annihilation of skyrmions involves rather complex winding and unwinding processes~\cite{2013_Milde_Science, 2017_Wild_SciAdv, 2017_Kagawa_NatCommun}. For instance, when undergoing the transition into the topologically trivial helical state, Bloch points initiate the coalescence of neighboring skyrmions. At low temperatures, this process may give rise to magnetic textures that share similarities with both a small-domain helical state incorporating topological disclination defects at the domain boundaries~\cite{2016_Dussaux_NatCommun, 2017_Bauer_PhysRevB, 2018_Schoenherr_NatPhys} and a state composed of elongated skyrmions~\cite{2017_Morikawa_NanoLett}.

Recently, in addition to the well-established high-temperature skyrmion lattice state (HTS) \cite{2012_Seki_Science, 2012_Adams_PhysRevLett}, a disconnected low-temperature skyrmion state~(LTS) was identified in Cu$_{2}$OSeO$_{3}$ for magnetic fields applied along $\langle100\rangle$ only, highlighting the crucial role of magnetocrystalline anisotropies for the stabilization of the LTS~\cite{2018_Chacon_NatPhys, 2014_Leonov_ArXiv, 2019_Bannenberg_npjQuantumMater}. The nucleation of skyrmions at low temperatures is facilitated by a tilted conical state as an intermediate state~\cite{2018_Qian_SciAdv, 2018_Halder_PhysRevB}. The resulting spin texture typically exhibits a high degree of disorder and a pronounced history dependence, most notably the volume fraction of LTS may be increased by magnetic field cycling~\cite{2018_Chacon_NatPhys, 2019_Mettus_unpublished}.

Although the suppression of finite-temperature effects initially limits the nucleation of the LTS, it also permits to address the intrinsic ground-state properties of topologically nontrivial objects, complementary to studies of the metastable HTS frozen-in by means of field cooling~\cite{2010_Munzer_PhysRevB, 2013_Ritz_PhysRevB, 2016_Bauer_PhysRevB, 2016_Oike_NatPhys, 2016_Karube_NatMater, 2016_Okamura_NatCommun, 2018_Bauer_Book, 2018_Chacon_NatPhys, 2018_Berruto_PhysRevLett, 2019_Bannenberg_npjQuantumMater, 2019_Birch_PhysRevB}. In this context, one aspect concerns the microwave dynamics, which in the cubic chiral magnets are well-understood for weak spin--orbit coupling~\cite{2012_Mochizuki_PhysRevLett, 2012_Onose_PhysRevLett, 2015_Schwarze_NatMater} and might be even exploited for new concepts in spintronic devices~\cite{2013_Nagaosa_NatNanotechnol, 2013_Okamura_NatCommun, 2015_Ogawa_SciRep, 2017_Garst_JPhysD}. The characteristic excitations comprise +Q and --Q modes in the helimagnetic phases as well as one breathing and two gyrational modes in the HTS. So far, however, no information was available on the microwave dynamics in the LTS and the tilted conical phase arising for larger spin--orbit coupling.

Using broadband ferromagnetic resonance measurements in combination with linear spin-wave theory, we show that the excitation spectra in the LTS are highly reminiscent of those in the HTS, despite distinctly different degrees of disorder and different stabilization mechanisms, namely thermal fluctuations for the HTS~\cite{2009_Muhlbauer_Science, 2013_Buhrandt_PhysRevB} and magnetocrystalline anisotropies for the LTS~\cite{2018_Chacon_NatPhys, 2014_Leonov_ArXiv}. All material-specific parameters entering the theory were determined in previous studies~\cite{2015_Schwarze_NatMater, 2018_Chacon_NatPhys, 2018_Halder_PhysRevB}, allowing for a parameter-free comparison between theory and experiment. By reducing the symmetry, the anisotropies mediate the hybridization of the breathing mode with a dark octupole gyration mode in the LTS. In addition, the excitations observed at low fields under decreasing field indicate the presence of a breathing mode not expected for the helimagnetic ground state, suggesting an oblique instability of the hexagonal skyrmion lattice driven by the elongation of skyrmions.

For the present study a single-crystal cuboid of dimensions $1.37\times1.5\times1.82~\mathrm{mm}^{3}$ oriented along $[001]$, $[110]$, and $[1\bar{1}0]$, was carefully polished and placed on a coplanar waveguide~(CPW) with one of the $(001)$ surfaces facing down. Two sample positions were measured; the sample was centered either on the central signal line (width: 1~mm) or on one of the gaps (width: 0.3~mm). This way, the exciting ac magnetic field was dominated by either in-plane or out-of-plane components, allowing to address selectively different resonant modes. As the lateral dimensions of sample and CPW were comparable, the excitation always contained a weak contribution of the nondominant component~\cite{2017_Stasinopoulos_SciRep}. If not stated otherwise, data shown in the following were recorded under in-plane excitation. The static magnetic field was applied parallel to $[001]$, i.e., perpendicular to the plane of the CPW, and was changed in steps of 1~mT. Low temperatures and static magnetic fields were provided by a flow cryostat with a superconducting magnet. All measurements were carried out at a temperature of 5~K. Using a vector network analyzer, at each field value the complex transmission $S_{21}$ through the CPW was measured while increasing the frequency $f$ from 1~GHz to 6.5~GHz in steps of 0.8~MHz. Background contributions were removed by subtracting a spectrum recorded at high magnetic field (500~mT), yielding the difference $\Delta S_{21}$ shown in the following. Complementary magnetization and ac susceptibility measurements were carried out using a Quantum Design physical properties measurement system.

\begin{figure}
	\includegraphics[width=0.8\linewidth]{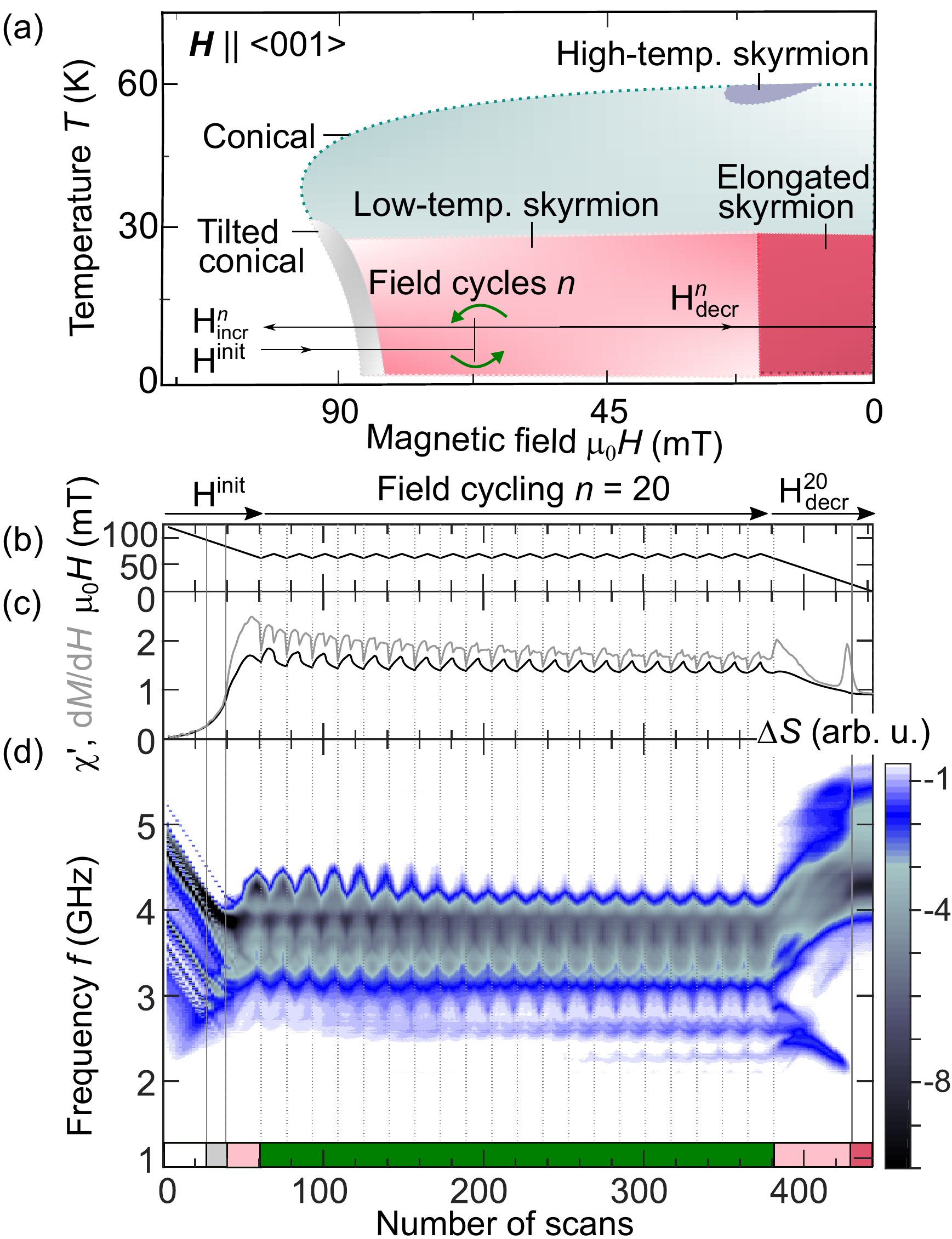}
	\caption{\label{fig:1}Evolution of magnetic properties under field cycling. (a)~Magnetic phase diagram as observed under decreasing field. Arrows illustrate the field histories used in the following. \mbox{(b,c)}~Magnetic field, $\mu_{0}H$, differential susceptibility, $\mathrm{d}M/\mathrm{d}H$, and real part of the ac susceptibility, $\chi'$. Data were recorded under decreasing magnetic field with $n = 20$ field cycles in the LTS. (d)~Microwave spectra. The color encodes the transmission difference $\Delta S_{21}$. With field cycling spectral weight is redistributed. The colored bar at the bottom indicates the dominating magnetic state, vertical solid lines indicate phase transitions. See text for details.}
\end{figure}

The magnetic phase diagram of Cu$_{2}$OSeO$_{3}$, shown in Fig.~\ref{fig:1}(a), represents an important point of reference for the discussion of the experimental data. As typical for a cubic chiral magnet, it comprises paramagnetic and field-polarized regimes at high temperatures and fields, respectively, the conical state described in terms of the superposition of a homogeneous magnetization and a helix both oriented along the field direction, and the high-temperature skyrmion lattice state in finite fields just below the transition temperature $T_{c} = 58$~K~\cite{2012_Seki_Science, 2012_Adams_PhysRevLett, 2012_Seki_PhysRevB, 2012_Seki_PhysRevBa, 2014_Omrani_PhysRevB, 2016_Qian_PhysRevB, 2016_Bauer_Book}. In addition, for magnetic field parallel to $\langle100\rangle$, in Cu$_{2}$OSeO$_{3}$ a separate low-temperature skyrmion state~(LTS) is observed in the vicinity of the upper critical field $\mu_{0}H_{c2} \approx 80$~mT~\cite{2018_Chacon_NatPhys}. Although the LTS represents the thermodynamic ground state, its nucleation is extremely slow due to the complexity of the topological winding involved. In fact, a two-step process is observed in which the metastable tilted conical state, characterized by conical helices tilting away from the field direction, is required as an intermediate state prior to the nucleation of the LTS~\cite{2018_Halder_PhysRevB, 2018_Qian_SciAdv}.

This complex nucleation process is also reflected in a pronounced dependence of the spin texture and the associated magnetic properties on the temperature and field history, where Fig.~\ref{fig:1} depicts the situation under decreasing magnetic field, also including metastable states. For the present study, it is therefore imperative to focus on distinct measurement protocols described in the following. Starting in a high magnetic field well above $H_{c2}$, the field is decreased until reaching the LTS, i.e., 70~mT for the given sample shape. This initial measurement from high fields is referred to as $\mathrm{H}^{\mathrm{init}}$ scan. Next, the magnetic field is cycled $n$ times between 70~mT and 62~mT, distinctly increasing the volume fraction of LTS as will be discussed below. After this cycling, the magnetic field is either decreased, referred to as $\mathrm{H}^{n}_{\mathrm{decr}}$ scan, or increased, referred to as $\mathrm{H}^{n}_{\mathrm{incr}}$ scan.

A typical measurement consisting of a $\mathrm{H}^{\mathrm{init}}$ scan, 20 field cycles, and a $\mathrm{H}^{20}_{\mathrm{decr}}$ scan is depicted in Figs.~\ref{fig:1}(b) through \ref{fig:1}(d). The abscissa corresponds to consecutive data points/scans that are separated by field steps of 1~mT, the actual magnetic field value is shown in Fig.~\ref{fig:1}(b). The color bar at the bottom indicates the prevailing magnetic state using the color coding introduced in Fig.~\ref{fig:1}(a), where green shading indicates data that were recorded during the cycling of the field. Solid vertical lines mark the phase boundaries as inferred conveniently from the differential susceptibility, $\mathrm{d}M/\mathrm{d}H$, and the real part of the ac susceptibility, $\chi'$, both shown in Fig.~\ref{fig:1}(c), where we refer to Ref.~\cite{2018_Halder_PhysRevB} for a comprehensive account. Field cycling enhances signatures related to the LTS, such as the pronounced maximum at its low-field boundary, which is interpreted as an increasing volume fraction of LTS in agreement with previous reports~\cite{2018_Chacon_NatPhys, 2019_Mettus_unpublished}.

Typical microwave spectroscopy data are depicted in Fig.~\ref{fig:1}(d). Here and in the following, they are shown in terms of the transmission difference $\Delta S_{21}$ recorded as a function of frequency at each field value. Dark colors indicate strong absorption due to the excitation of resonant modes in the sample. We start the description at high magnetic fields (left), where a linear slope is characteristic for a Kittel resonance in the field-polarized regime. In addition to the prominent Kittel mode, at lower frequencies several standing spin wave modes are excited due to the low magnetic damping, $\alpha \approx 10^{-4}$, of Cu$_{2}$OSeO$_{3}$ at low temperatures~\cite{2017_Stasinopoulos_ApplPhysLett, 2017_Weiler_PhysRevLett}. These modes are not of central interest for the present study but nevertheless complicate the interpretation of the results at lower fields. As a consequence, the following discussion focuses on the dominant modes in each magnetic phase, for instance refraining from an analysis of the low-intensity clockwise gyration mode in the skyrmion phase.

When ignoring the field cycling for a moment, the microwave spectrum reminds of the well-established universal spectrum of the cubic chiral magnet~\cite{1977_Date_JPhysSocJpn, 2012_Onose_PhysRevLett, 2012_Koralek_PhysRevLett, 2015_Schwarze_NatMater, 2019_Pollath_PhysRevLett}, although it is more complex due to the presence of the tilted conical state and the LTS. Under decreasing magnetic field (left to right), a change of slope in the Kittel mode marks the onset of the tilted conical state. Decreasing the field further, a broad band of absorption increases in frequency, reminiscent of the behavior in the conical state. The field cycling is reflected in a zigzag shape and a continuous shift of spectral weight from high to low frequencies, in particular resulting in the emergence of a set of low-frequency modes. These modes essentially decrease in frequency under decreasing field until vanishing at the low-field boundary of the LTS.

\begin{figure}
	\includegraphics[width=0.8\linewidth]{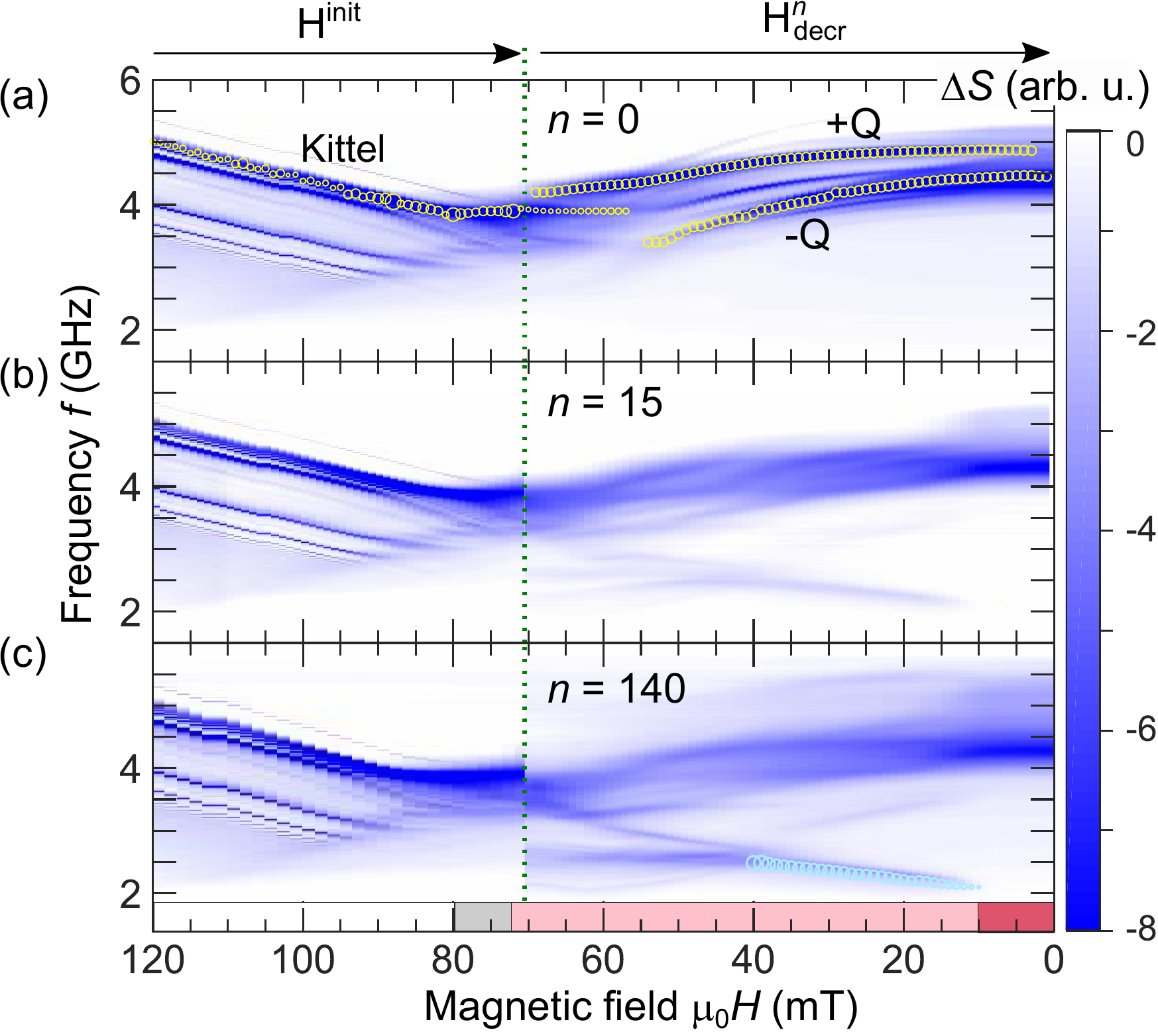}
	\caption{\label{fig:2}Evolution of the microwave spectra under field cycling. For each panel, the transmission difference $\Delta S_{21}$ was recorded during an initial $\mathrm{H}^{\mathrm{init}}$ scan to 70~mT (dotted green line), followed by $n$ field cycles (not shown), before finally decreasing the field to zero in a $\mathrm{H}^{n}_{\mathrm{decr}}$ scan. (a)~Without cycling the spectra are dominated by the $\pm$Q modes of the tilted conical and conical state (yellow circles). \mbox{(b,c)}~With increasing number of cycles spectral weight is continuously redistributed from the $\pm$Q modes to a previously unresolved set of low-frequency modes associated with the LTS (cyan circles).}
\end{figure}

The nature of the different modes is most efficiently discussed when also considering their evolution under field cycling, as elaborated on in Fig.~\ref{fig:2}, where initial $\mathrm{H}^{\mathrm{init}}$ scans are combined with $\mathrm{H}^{n}_{\mathrm{decr}}$ scans measured after different numbers, $n$, of field cycles. For clarity, the field cycles are omitted from the microwave spectra. In the $\mathrm{H}^{0}_{\mathrm{decr}}$ scan without cycling, shown in Fig.~\ref{fig:2}(a), two distinct modes may be distinguished that essentially increase in frequency for decreasing field. These modes are marked by gray circles and remind of the +Q and --Q modes of the conical state. Additional absorption at reduced spectral weight follows a qualitatively similar field dependence, suggesting standing spin wave modes as its origin. Changes of slope and a small discontinuity in the $\pm$Q modes near $H_{c2}$ are attributed to the tilted conical state, in which the finite angle between field and propagation direction induces small deviations in the microwave response, akin to those in the helical state at small fields, cf.\ Supplementary Information of Ref.~\cite{2015_Schwarze_NatMater}. 

Moderate cycling, such as in the $\mathrm{H}^{15}_{\mathrm{decr}}$ scan shown in Fig.~\ref{fig:2}(b), reduces the spectral weight of the $\pm$Q modes at intermediate fields. Instead, a set of modes emerges at distinctly lower frequencies of about 2.5~GHz. Increasing the number of cycles to $n = 140$, as shown in Fig.~\ref{fig:2}(c), intensifies this shift of spectral weight. The emerging modes exhibit frequencies and an evolution under decreasing field that are reminiscent of the counterclockwise gyration mode in the high-temperature skyrmion lattice. Also taking into account the redistribution of spectral weight under field cycling and the disappearance of the modes at the low-field boundary of the LTS, these findings consistently suggest that the low-frequency modes are associated with the LTS. From Lorentzian fits, narrow line widths are extracted, translating to small damping constants $\alpha \le 0.01$ for these skyrmion modes. Note that the observation of resonant modes characteristic of skyrmions also provides direct evidence for the fixed phase relationship of the multi-$Q$ structure underlying the nontrivial topology of the LTS~\cite{2011_Adams_PhysRevLett, 2019_Kindervater_PhysRevX}.

\begin{figure}
	\includegraphics[width=0.8\linewidth]{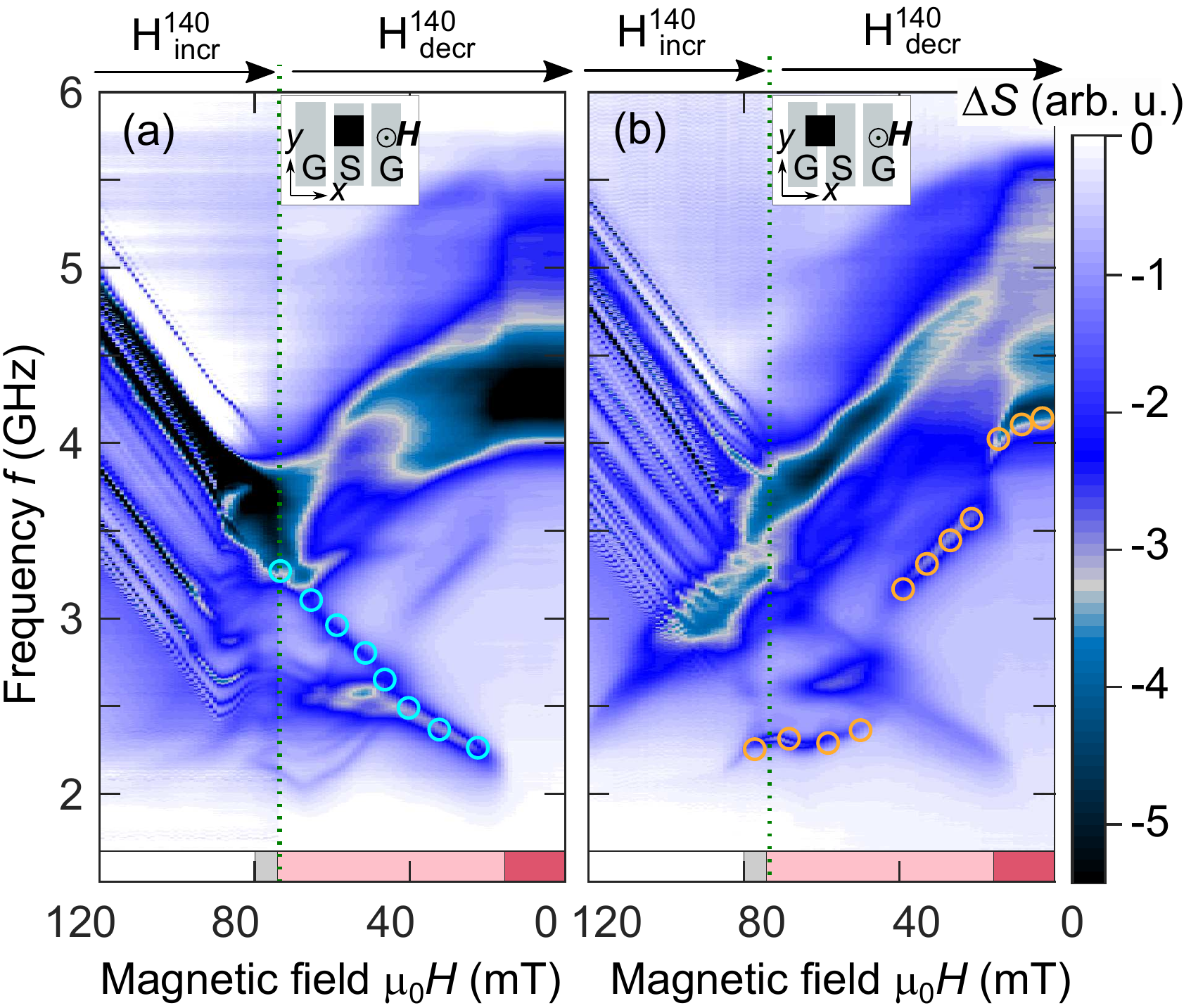}
	\caption{\label{fig:3}Microwave spectra after field cycling for different excitation geometries as schematically depicted in the insets. Each panel is composed of two independent scans both obtained after 140 field cycles, i.e., an $\mathrm{H}^{140}_{\mathrm{incr}}$ and an $\mathrm{H}^{140}_{\mathrm{decr}}$ scan. (a)~Sample centered on the signal line of the CPW yielding predominantly in-plane~($h_{x}$) excitation. (b)~Sample centered on one of the gaps of the CPW yielding predominantly out-of-plane~($h_{z}$) excitation. Circles mark the characteristic modes in the LTS, namely the counterclockwise gyration mode (cyan) and the breathing mode (orange).}
\end{figure}

Further information on the character of the different modes, in particular those in the LTS, is obtained from measurements under different excitation geometries as shown in Fig.~\ref{fig:3}. In order to cover the entire field range of interest, each panel in Fig.~\ref{fig:3} combines data from an $\mathrm{H}^{140}_{\mathrm{incr}}$ and an $\mathrm{H}^{140}_{\mathrm{decr}}$ scan. By placing the sample either on the center of the signal line or on one of the gaps of the CPW, the microwave excitation in the sample is dominated either by in-plane or out-of-plane components. In skyrmion states, in-plane excitation couples efficiently to gyration modes, while out-of-plane excitation drives breathing modes~\cite{2012_Mochizuki_PhysRevLett}. In helical or conical states, only excitation components perpendicular to the propagation vector couple to the $\pm$Q modes~\cite{2012_Onose_PhysRevLett}. Consequently, modes observed under out-of-plane excitation are associated with magnetic structures that enclose finite angles with the magnetic field, such as multi-domain helical or tilted conical states. 

Under in-plane excitation, as shown in Fig.~\ref{fig:3}(a) and all figures discussed so far, two groups of modes are prevailing below $H_{c2}$; (i) the modes in the conical and tilted conical state that increase in frequency with decreasing field, and (ii) a lower-frequency mode that decreases with decreasing field and is associated with the LTS (marked by cyan circles). When out-of-plane excitation dominates, as shown in Fig.~\ref{fig:3}(b), the overall spectral weight is reduced with weak remnants of the previously discussed modes being excited due to small in-plane excitation components. In addition, a prominent broad mode in the frequency range of the $\pm$Q modes is associated with the tilted conical state. Perhaps most intriguingly, however, a distinct low-frequency mode is observed across the entire field range below $H_{c2}$ (marked by orange circles).

As substantiated by the theoretical analysis presented below, the sensitivity with respect to the excitation geometry identifies the modes marked by cyan and orange circles as the counterclockwise gyration mode and the breathing mode in the LTS. As a function of decreasing field, the resonance frequency of the breathing mode exhibits two abrupt increases. In the following we will establish that the jump at ${\sim}$40~mT arises from anti-crossing due to a hybridization with a dark octupole gyration mode that is also observed in the quenched high-temperature skyrmion lattice~\cite{2019_Seki_privatecommunication}, while the jump at ${\sim}$20~mT is consistent with the formation of elongated skyrmions.

\begin{figure}
	\includegraphics[width=0.8\linewidth]{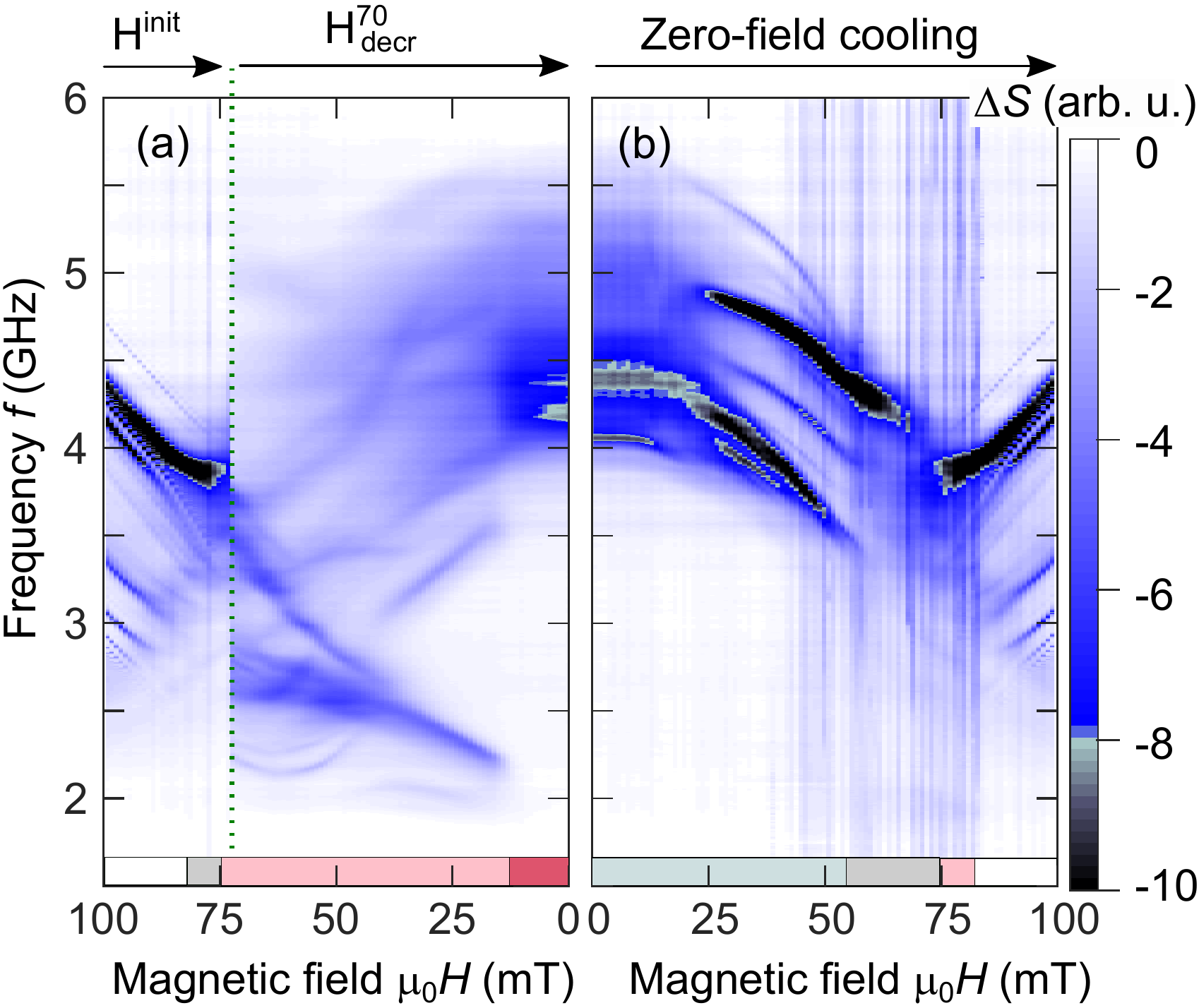}
	\caption{\label{fig:4}Comparison of microwave spectra after field cycling and after initial zero-field cooling. (a)~Data for decreasing magnetic field after field cycling composed of an $\mathrm{H}^{\mathrm{init}}$ and an $\mathrm{H}^{70}_{\mathrm{decr}}$ scan. (b)~Data for increasing magnetic field after initial zero-field cooling. Note in particular the discrepancy of the resonance frequencies in zero field. See text for details.}
\end{figure}

The presence of elongated skyrmions is supported by a comparison of microwave spectra recorded after field cycling, i.e., a $\mathrm{H}^{70}_{\mathrm{decr}}$ scan, with corresponding spectra recorded after initial zero-field cooling in Fig.~\ref{fig:4}. Note that the latter data are measured under increasing magnetic field, leading to a modified sequence of phase transitions, where we refer to Ref.~\cite{2018_Halder_PhysRevB} for a detailed account on this hysteresis. After zero-field cooling, a multi-domain helical state forms in Cu$_{2}$OSeO$_{3}$ with three equally populated domains of helices oriented along the $\langle100\rangle$ axes, representing the thermodynamic ground state. A magnetic field pointing along one of the $\langle100\rangle$ axes favors the domain aligned with the field, resulting in an increase of its population upon increasing field and, eventually, in a single-domain conical state. This process is practically irreversible at low temperatures and thus a single helimagnetic domain is obtained after removing the field~\cite{2010_Bauer_PhysRevB, 2017_Bauer_PhysRevB}. When decreasing the field starting from the LTS, the situation is markedly different and the system appears to be trapped in a metastable state.

As shown in the central part of Fig.~\ref{fig:4}, the resonance frequencies observed in zero field after field cycling, cf.\ Fig.~\ref{fig:4}(a), differ decisively from those observed in the well-ordered multi-domain helical state after zero-field cooling, cf.\ Fig.~\ref{fig:4}(b). This discrepancy suggests that the dynamic properties of the complex zero-field magnetic texture obtained after field cycling are not captured correctly by a description in terms of helical domains. Instead, the calculations presented in the following imply that, at least from the point of view of microwave excitations, a description in terms of elongated skyrmions is more accurate.

Our theoretical treatment follows previous work~\cite{2015_Schwarze_NatMater} and uses the standard phenomenological model for chiral magnets supplemented by magnetocrystalline anisotropies. This anisotropy contribution to the free energy, $\varepsilon_{a}$, proves to be key for the description of the properties of the LTS in Cu$_{2}$OSeO$_{3}$. The space group $P2_{1}3$ allows various magnetocrystalline terms but we demonstrate that already the simplest form, $\varepsilon_{a} = K (m_{x}^{4} + m_{y}^{4} + m_{z}^{4})$, captures the fundamental aspects when using the anisotropy constant $K = -2\cdot10^{3}~\mathrm{J}/\mathrm{m}^{3}$ as in Ref.~\cite{2018_Chacon_NatPhys}, cf.\ Supplemental Material~\cite{} for considerations on other values of $K$. In order to address the resonances of all experimentally observed magnetic textures across the entire field range, excitation frequencies were determined for (i)~non-modulated and one-dimensionally modulated states, i.e, the field-polarized, tilted conical, and conical states, within their respective (meta-)stability range, and (ii)~two-dimensionally modulated states, i.e., the topologically nontrivial skyrmion states. In these calculations, the $\pm$Q modes and the gyrational modes are driven by in-plane excitation, while the breathing mode is driven by out-of-plane excitation. The resonance in the tilted conical state is excited in both cases.

\begin{figure}
	\includegraphics[width=0.8\linewidth]{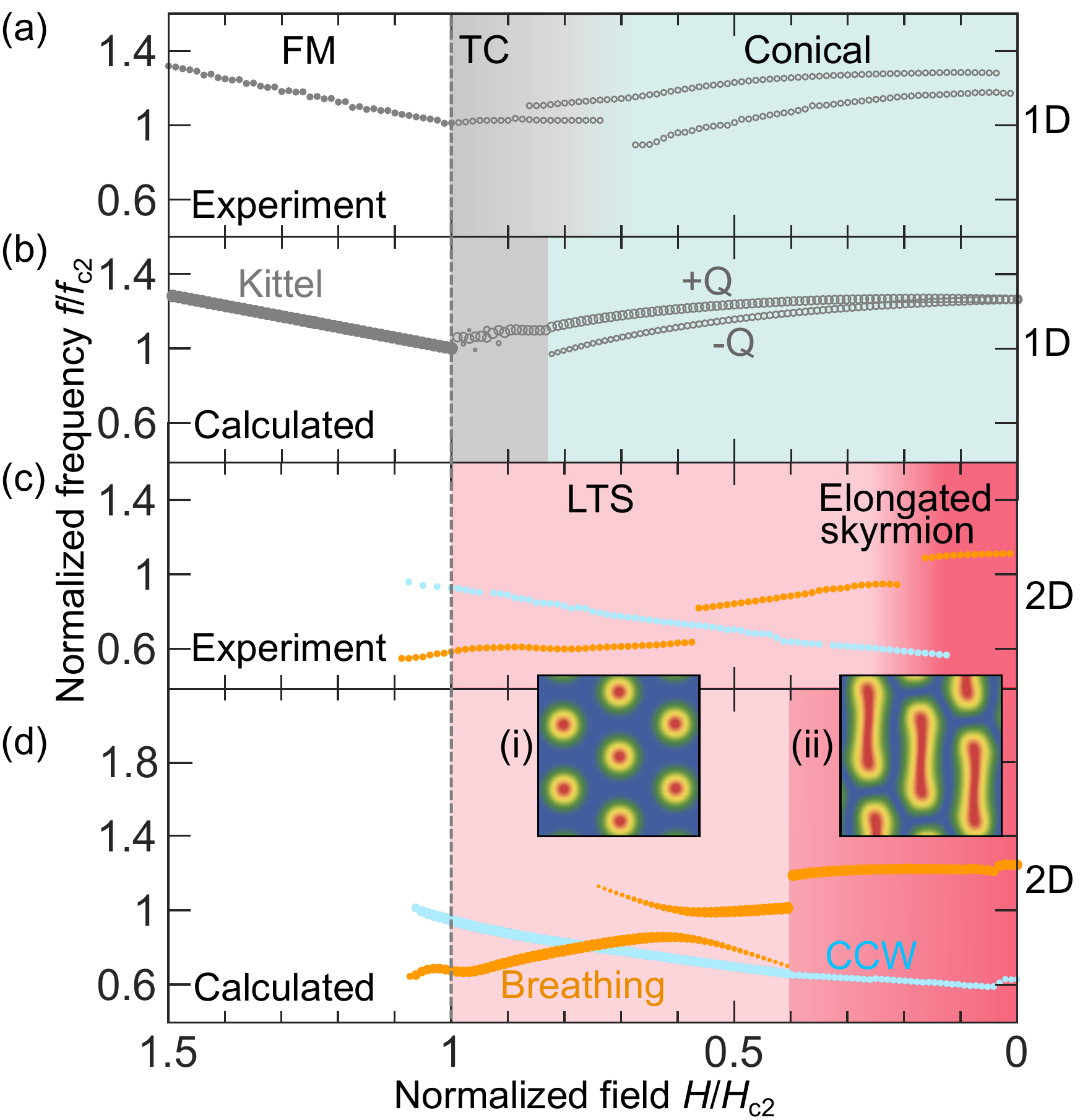}
	\caption{\label{fig:5}Comparison of experimental and calculated spectra. Frequency and magnetic field values are normalized to their respective values at $H_{c2}$. (a)~Experimental spectra for the field-polarized, tilted conical, and conical states. Data inferred from Fig.~\ref{fig:2}(a). (b)~Calculated spectra for the non-modulated and one-dimensionally modulated states. (c)~Experimental spectra for the skyrmion states. Data inferred from Fig.~\ref{fig:4}. (d)~Calculated spectra for the two-dimensionally modulated states. A counterclockwise gyration mode (cyan symbols) and a breathing mode (orange symbols) are identified. Symbol sizes in (b) and (d) represent spectral weight. \mbox{(i,ii)}~Calculated real-space images of the magnetic texture associated with the breathing mode at high and low fields.}
\end{figure}

For $K = 0$ (not shown) the calculations reproduce the characteristic modes as observed in the conical and the skyrmion lattice state at high temperatures where $K$ is small~\cite{2012_Onose_PhysRevLett, 2015_Schwarze_NatMater}. As shown in the experimental spectra in Fig.~\ref{fig:5}(a) and the calculated spectra in Fig.~\ref{fig:5}(b), a finite anisotropy leaves the Kittel mode in the field-polarized regime and the $\pm$Q modes in the conical state highly reminiscent of their counterparts in the isotropic case. At magnetic fields just below $H_{c2}$, however, the tilted conical state (gray shading) becomes energetically more favorable than the regular conical state, as reflected by a change of slope, kinks, and minor discontinuities in the +Q mode. Note that the irregular field dependence of the resonance frequency originates in fact in a multitude of hybridizations.

In both experimental and calculated spectra of the LTS, shown in Figs.~\ref{fig:5}(c) and \ref{fig:5}(d), the counterclockwise gyration mode (cyan symbols) dominates under in-plane excitation. Finite anisotropy leaves the character of this mode essentially unchanged, i.e., its frequency monotonically decreases with decreasing field akin to the counterclockwise gyration mode in the high-temperature skyrmion lattice. In contrast, the breathing mode (orange symbols) prevailing in the LTS under out-of-plane excitation is subject to fundamental changes, with good agreement between experiment and theory. While for the isotropic case, $K = 0$, the frequency of the breathing mode monotonically increases with decreasing field, for $K \neq 0$ a small local minimum just below $H_{c2}$ is followed by characteristic anti-crossing around $0.6\,H_{c2}$. This signature is the hallmark of the hybridization with a dark octupole gyration mode mediated by the magnetocrystalline anisotropies. At lower fields, around $0.4\,H_{c2}$, the hexagonal skyrmion lattice becomes unstable with respect to an oblique distortion. This instability is driven by the elongation of skyrmions as illustrated by the real-space images in Figs.~\ref{fig:5}(i) and \ref{fig:5}(ii). The frequency of the breathing mode is enhanced for the elongated case, cf.\ Supplemental Material~\cite{} for animations, and comes close to that of $\pm$Q modes of the helimagnetic order but remains distinctly lower, in agreement with the experimental data shown in Fig.~\ref{fig:4}. On a similar note, the spectral weight of the counterclockwise gyration mode is distinctly reduced for the elongated skyrmions, consistent with its disappearance in the experimental spectra.

The oblique instability of the skyrmion lattice occurs in a field range where skyrmions are only metastable. Similar to metastable isolated skyrmions~\cite{1994_Bogdanov_PhysStatusSolidiB}, they experience an elliptical instability and become elongated, leading to an oblique distortion of the hexagonal lattice. As topological unwinding is energetically costly, the history employed in Fig.~\ref{fig:5}(c) favors the observation of this oblique instability, notably the magnetic field is decreased at low temperatures after the nucleation of the topologically nontrivial LTS by means of field cycling. Following this history, the observation of a low-field resonance that is, in contrast to the helical $\pm$Q modes, susceptible to out-of-plane excitation (orange symbols) suggests that a (metastable) oblique skyrmion state hosting elongated skyrmions was indeed realized in the experiment.

In conclusion, the cubic chiral magnet Cu$_{2}$OSeO$_{3}$ was studied by means of broadband ferromagnetic resonance measurements, focusing in particular on the low-temperature skyrmion state. Employing field cycling, two excitation geometries, and comparing the experimental results to linear spin-wave theory, resonant modes in the conical, tilted conical, and low-temperature skyrmion state are clearly identified, with the breathing mode in the LTS exhibiting a characteristic hybridization. Most intriguingly, the resonances observed in small fields after field cycling indicate the presence of elongated skyrmions. These findings not only highlight how the robustness of topological non-triviality may influence dynamic properties of magnetic materials, but also showcase how the study of dynamic properties may provide valuable insights to static properties, such as the microscopic nature of magnetic textures.

We wish to thank G.~Benka, A.~Chacon, and S.~Mayr for fruitful discussions and assistance with the experiments. This work has been funded by the Deutsche Forschungsgemeinschaft (DFG, German Research Foundation) under SPP2137 (Skyrmionics, Project No.\ 360506545, Projects 403030645, 403191981, and 403194850), TRR80 (From Electronic Correlations to Functionality, Project No.\ 107745057, Projects E1, F7, and G9), and the excellence cluster MCQST under Germany's Excellence Strategy EXC-2111 (Project No.\ 390814868). This project has received funding from the European Metrology Programme for Innovation and Research (EMPIR) programme co-financed by the Participating States and from the European Union’s Horizon 2020 research and innovation programme. A.B., D.M., and C.P.\ acknowledge financial support through the European Research Council (ERC) through Advanced Grant No.\ 788031 (ExQuiSid). T.T.\ acknowledges funding by the JSPS Overseas Research Fellowship. M.G.\ is supported by DFG SFB1143 (Correlated Magnetism: From Frustration To Topology, Project No.\ 247310070, Project A07), DFG Grant No.\ 1072/5-1 (Project No.\ 270344603), and DFG Grant No.\ 1072/6-1 (Project No.\ 324327023).

%

\end{document}